\documentclass[3p,preprint,11pt,letterpaper]{elsarticle}
\pdfoutput=1 
\usepackage{calc}
\usepackage{amsmath}
\usepackage{amssymb}
\usepackage{amsmath}
\usepackage{textcomp}
\usepackage{color}
\usepackage{framed}
\usepackage{ctable}
\usepackage{tabularx}
\usepackage{hyperref}
\usepackage{wrapfig}
\usepackage{xspace}
\usepackage{multirow}

\hypersetup{colorlinks=true,breaklinks=true,linkcolor=blue,urlcolor=blue,citecolor=blue}
\definecolor{shadecolor}{rgb}{0.9,0.9,0.9}

\journal{Computer Physics Communications}

\usepackage{graphicx}
\usepackage{amssymb}

\usepackage{lineno}

\definecolor{darkblue}{rgb}{0,0,.6}
\definecolor{darkred}{rgb}{.6,0,0}
\definecolor{darkgreen}{rgb}{0,.6,0}
\definecolor{red}{rgb}{.98,0,0}

\def\ssmall{\fontsize{8pt}{2pt}\selectfont}

\usepackage{listings}
\usepackage{caption}

\DeclareCaptionFormat{listing}{\rule{\dimexpr\textwidth+17pt\relax}{0.4pt}\par\vskip1pt#1:#2#3}
\lstnewenvironment{cpplisting}[1][]
  {\lstset{language=C++,numbers=right,#1}}{}

\lstnewenvironment{bashlisting}[1][]
  {\lstset{language=bash,numbers=right,#1}}{}

\lstnewenvironment{outputlisting}[1][]
  {\lstset{language=html,numbers=right,#1}}{}

\newcommand{\code}[1]{\texttt{#1}}

\newcommand{\Boost}{\code{Boost}\xspace}
\newcommand{\MPI}{\code{MPI}\xspace}
\newcommand{\Doxygen}{\code{Doxygen}\xspace}
\newcommand{\CXX}{\code{C++}\xspace}
\newcommand{\Python}{\code{Python}\xspace}
\newcommand{\CMake}{\code{CMake}\xspace}
\newcommand{\GCC}{\code{GCC}\xspace}
\newcommand{\HDF}{\code{HDF5}\xspace}

\newcounter{bla}

\newcommand\bigO{{\mathcal O}}

\begin{document}

\begin{frontmatter}

\title{Updated Core Libraries of the ALPS Project}


\author[umich]{Markus~Wallerberger\corref{author}} \ead{mwallerb@umich.edu}
\author[umich]{Sergei~Iskakov}


\author[umich]{Alexander~Gaenko}
\author[umich]{Joseph~Kleinhenz}
\author[umich]{Igor~Krivenko}
\author[uiuc]{Ryan~Levy}
\author[umich]{Jia~Li}
\author[saitama]{Hiroshi~Shinaoka}
\author[utokyo2,utokyo3]{Synge~Todo}             

\author[wchesterpenn]{Tianran~Chen}
\author[simons]{Xi~Chen}         
\author[mun]{James~P.~F.~LeBlanc} 
\author[umich]{Joseph~E.~Paki}             
\author[tenn]{Hanna~Terletska}
\author[eth]{Matthias~Troyer}             

\author[umich]{Emanuel~Gull}
\cortext[author]{Corresponding author.}

\address[umich]{University of Michigan, Ann Arbor, Michigan 48109, USA}
\address[uiuc]{University of Illinois, Urbana-Champaign, Illinois 61820, USA}
\address[saitama]{Department of Physics, Saitama University, Saitama 338-8570, Japan}
\address[utokyo2]{Department of Physics, University of Tokyo, Tokyo 113-0033, Japan}
\address[utokyo3]{Institute for Solid State Physics, University of Tokyo, Kashiwa 277-8581, Japan}
\address[wchesterpenn]{West Chester University of Pennsylvania, West Chester, Pennsylvania 19383, USA}
\address[simons]{Center for  Computational Quantum Physics, Flatiron Institute, New York, NY 10010, USA}
\address[mun]{Department of Physics and Physical Oceanography, Memorial University of Newfoundland, St. John’s, Newfoundland \& Labrador A1B 3X7, Canada}
\address[tenn]{Middle Tennessee State University, Murfreesboro, Tennessee 37132, USA}
\address[eth]{Institut f\"ur Theoretische Physik, ETH Zurich, 8093 Zurich, Switzerland}

\begin{abstract}
The open source ALPS (Algorithms and Libraries for Physics Simulations) project 
provides a collection of physics libraries and applications, with a focus on 
simulations of lattice models and 
strongly correlated electron systems. The libraries provide a convenient set of 
well-documented and reusable components for developing condensed matter physics 
simulation codes, and the applications strive to make commonly used and proven 
computational algorithms available to a non-expert community. In this paper we 
present an update of the core ALPS libraries. We present in particular new 
Monte Carlo libraries and new Green's function libraries.
\end{abstract}

\end{frontmatter}



\noindent {\bf PROGRAM SUMMARY}

\begin{small}
\noindent
{\em Program Title:} ALPS Core libraries 2.3        \\
{\em Project homepage:} http://alpscore.org                \\
{\em Catalogue identifier:} --                                  \\
{\em Journal Reference:} --                                     \\
{\em Operating system:} Unix, Linux, macOS \\
{\em Programming language:} \verb*#C++11# or later\\
{\em Computers:}
  any architecture with suitable compilers including PCs, clusters and
  supercomputers\\
{\em Licensing provisions:} GNU General Public License (GPLv2+)\\
{\em Classification:}
6.5, 
7.3, %
20 
\\
{\em Dependencies:} \CMake ($\ge 3.1$), 
  \Boost ($\ge 1.56$), 
  \HDF ($\ge 1.8$),
  {\tt Eigen} ($\ge 3.3.4$)\\
{\em Optional dependencies:}
   \MPI ($\ge 2.1$), {\tt Doxygen}
\\
{\em Nature of problem:}
   Modern, lightweight, tested and documented libraries covering
   the basic requirements of rapid development of efficient physics
   simulation codes, especially for modeling strongly correlated electron systems.
\\
{\em Solution method:}
   We present \CXX open source computational libraries that
provide a convenient set of components for developing parallelized
physics simulation codes. The library features a short development
cycle and up-to-date user documentation.
\end{small}

\section{Introduction}
\label{sec:introduction}
The open source ALPS (Algorithms and Libraries for Physics Simulations) project \cite{ALPSCore17,ALPS10,ALPS13,ALPS2.0} provides a collection of computational physics libraries and applications, especially suited for the simulation of lattice models and strongly correlated systems.  The libraries are lightweight, thoroughly tested and documented.  The libraries are targeted at code developers who implement generic algorithms and utilities for rapid development of efficient computational physics applications.

Computer codes based on the ALPS (core) libraries have provided physical insights into many subfields of condensed matter. Highlights include nonequilibrium dynamics \cite{Eckstein09,Werner09}, continuous-time quantum Monte Carlo \cite{Werner06,Gull08,Shinaoka:2017cpa,CTINT}, LDA+DMFT materials simulations \cite{Ferber12}, simulations of quantum \cite{Cremades12} and classical \cite{Bergman07} spin systems, correlated boson \cite{Pollet10} and fermion \cite{LeBlanc15} models, as well as cuprate superconductivity \cite{Gull13}.

In this paper we present an updated and refactored version of the ALPS core libraries.  This version introduces a new module to the library: ALPS ALEA, which is designed as a replacement of the previous accumulator framework.  It provides significant runtime speed-ups for all of its accumulator types, improved compile times, as well as support for complex random variables, non-linear propogation of uncertainty and statistical hypothesis testing as a way to compare stochastic results (Section~\ref{sec:alea}).

Improvements and bug fixes were made to the existing modules.  The Green's function library was rewritten to be faster and more versatile (Section~\ref{sec:gf}).  Section~\ref{sec:components} briefly discusses the other components of the library with the corresponding improvements.

An effort was made to make the installation procedure more robust and straight-forward. In Section~\ref{sec:prereqs}, we list the software packages prerequisite for compiling and installing the library, as well as outline the installation procedure.  In  Section~\ref{sec:cite} we specify the ALPS citation policy and license, and in Sections~\ref{sec:summary} and~\ref{sec:ackn} we summarize the paper and acknowledge our contributors. Additionally, \ref{sec:installation-detail} provides installation examples.

\section{ALEA library}
\label{sec:alea}

The ALEA library (namespace \code{alps::alea}) is the next-generation statistics
library for Markov chain Monte Carlo calculations in the ALPS core libraries.
It is intended to supersede the old accumulators library
(namespace \code{alps::accumulators}), which is still supported for backwards
compatibility.  Compared to the old library, ALEA features:

\begin{itemize}
 \item improved runtime scaling for full binning and logarithmic binning
       accumulators, which were both reduced from $\bigO(N\log N)$ to $\bigO(N)$ in
       terms of the number of samples $N$.\cite{Wallerberger18}

 \item improved runtime performance for vector-valued accumulators due to the
       use of the Eigen library and the avoidance of the creation of temporary
       objects.

 \item improved memory footprint: (a) the use of \code{finalize()} semantics
       avoids duplicating the accumulator's data when evaluating the mean
       and variance; (b) \code{computed} objects allow creating and adding
       a new measurement in one step without the need to create a temporary.

 \item improved full binning procedure allows the retention of more fine-grained
       data while keeping linear runtime scaling.\cite{Wallerberger18}

 \item improved compilation times and error messages: (a) the interface was
       greatly simplified, the metaprogramming logic was removed, and the numerics
       are now in the compiled library instead of the headers; (b) the compile-time
       dependence on MPI and HDF5 was removed and superseded by clearly specified
       interfaces.

 \item native support for complex random variables, with user choice of
       computing either circular error bars or an error ellipse.

 \item support for computing the full covariance matrix.

 \item support for statistical hypothesis testing \cite{Wallerberger17} as a
       robust way to compare Monte Carlo results.
\end{itemize}

\begin{figure}
  \begin{center}
      \includegraphics[width=.5\columnwidth]{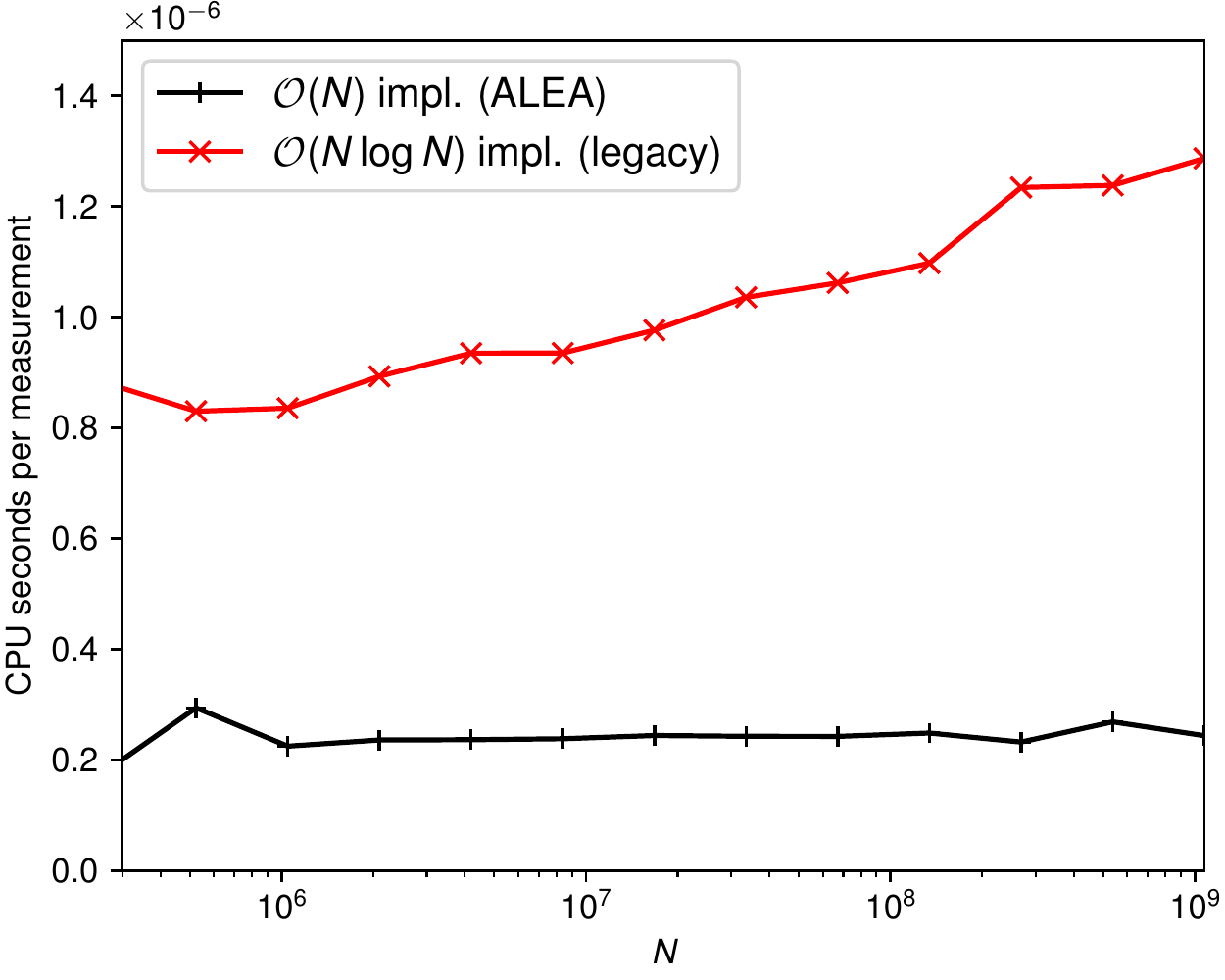}
  \end{center}
  \caption{CPU time spent per data point when accumulating $N$ data points of
           a random vector with $k=100$ components for the legacy accumulators
           framework (red crosses) and the ALEA library (black plusses).}
  \label{fig:cputime}
\end{figure}

To illustrate the improvements, we compare the CPU time needed per data point for
accumulating the autocorrelation time for a random vector with $k=100$ components
between the old accumulators framework and ALEA (Figure~\ref{fig:cputime}).  One
can see that there is both an offset to the scaling, which comes from avoiding
temporaries, as well as a better overall $\bigO(N)$ vs. $\bigO(N\log N)$ scaling.

\paragraph{Accumulators}
In Monte Carlo simulations, we are interested in obtaining statistical
properties of some random vector $X$ with $k$ components by estimating
it from a sequence of $N$ samples or ``measurements'' $x=(x_{1},x_{2}\ldots,x_{N})$.
Typically, $N$ is so large that retaining the individual sample points
is impossible or at least impractical. ALEA accumulators provide a workaround
by retaining only certain aggregates of the measurements by performing
reductions over the measurements on the fly, one measurement at a
time.

ALEA defines five accumulators, which differ in the stored
statistical estimates and associated runtime and memory cost (cf. Table~\ref{tab:acc}).
They represent commonly used cases ranging from fast and memory efficient to slow
but precise. In all accumulators, \code{size()} returns $k$, while \code{count()}
returns $N$.
\begin{table}
\begin{centering}
\begin{tabular}{llllccccc}
\toprule
ALEA name & old name & time & memory & \code{mean} & \code{var} & \code{cov} & \code{tau} & \code{batch}\tabularnewline
 &  & &  & $\langle x\rangle$ & $\sigma_{x}^{2}$ & $\Sigma_{x}$ & $\tau_{\mathrm{int},x}$ & $\bar{x}_{i}$\tabularnewline
\midrule
\code{mean\_acc} & \code{MeanAccumulator} & $Nk$ & $k$ & \checkmark &  &  &  & \tabularnewline
\code{var\_acc} & \code{NoBinningAccumulator} & $Nk$ & $k$ & \checkmark & \checkmark &  &  &
\tabularnewline
\code{cov\_acc} & --- & $Nk^{2}$ & $k^{2}$ & \checkmark & \checkmark & \checkmark &  & \tabularnewline
\code{autocorr\_acc} &  \code{LogBinningAccumulator} & $Nk$ & $k\log N$ & \checkmark & \checkmark
&  & \checkmark & \tabularnewline
\code{batch\_acc} & \code{FullBinningAccumulator} & $Nk$ & $Bk$ & \checkmark & \checkmark & \checkmark &
& \checkmark \tabularnewline
\bottomrule
\end{tabular}
\par\end{centering}
\caption{Accumulators in ALEA. Asymptotic complexity $\bigO(\cdots)$ for memory
and runtime in terms of sample size $N$, number of components of random
vector $k$, and number of bins $B$. Available statistical estimates
are marked with ``\checkmark'', where $\langle x\rangle$ is the sample mean, $\sigma_{x}^{2}$ is the sample variance,
$\Sigma_{x}$ is the sample covariance matrix, $\tau_{\mathrm{int},x}$ is the logarithmic binning estimate
for the integrated autocorrelation time, and $\bar{x}_{i}$ are bin means. }
\label{tab:acc}
\end{table}

\paragraph{Complex variances}
In the case of complex random variables, the concept of error bars becomes
somewhat ambiguous. ALEA supports two common strategies by admitting an additional
\code{Str} template argument to \code{var\_acc} and \code{cov\_acc}:

\begin{itemize}
\item Error with circularity constraint (\code{circular\_var}): error based on
the distance from the mean (circle) in the complex plane.  Provides an upper bound to
the total error and ignores any error structure in the real and imaginary part
of the variable.
\item Error ellipse (\code{elliptic\_var}): Treats the real and imaginary
part of a complex variable as two separate random variables, thus creating an
error ellipse around the mean in the complex plane.  Retains the full error
structure in the complex plane and allows one to plot separate error bars for
real and imaginary part.
\end{itemize}

\paragraph{Accumulators and results}
\begin{figure}
\begin{centering}
\includegraphics[scale=1.15]{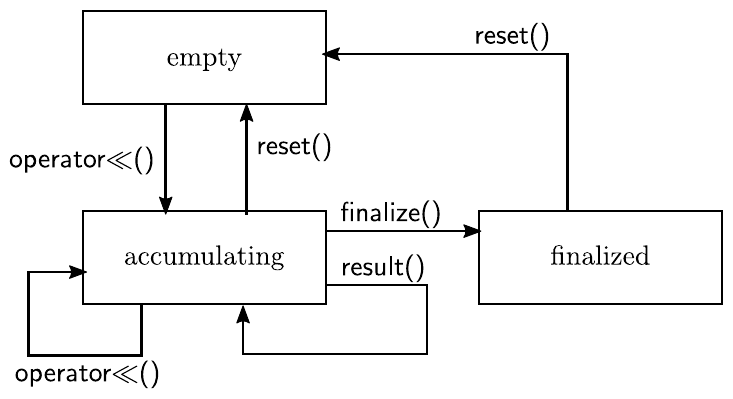}
\par\end{centering}
\caption{Accumulator and result types: finite state machine for accumulator
lifecycle.  The boxes ``empty'', ``accumulating'' and ``finalized'' represent
the possible states of the accumulator.  Calling the method indicated on the
edges takes the accumulator from one state to another.}
\label{fig:state}
\end{figure}

Each accumulator type has a matching result type (e.g., \code{mean\_acc} has
a matching \code{mean\_result}).  Accumulators and results are facade types
over a common base type (e.g., \code{mean\_data}), but differ conceptually
and thus have complementary functionality: accumulators support adding data
to them, while results allow one to perform transformations, reductions,
as well as extracting statistical estimates such as the sample mean.

To obtain a result from an accumulator, the accumulators provide both
a \code{result()} and a \code{finalize()} method. The \code{result()}
method creates an intermediate result, which leaves the accumulator untouched
and thus must involve a copy of the data, while the \code{finalize()} method
invalidates the accumulator and repurposes its internal data as simulation 
result
(it replaces the sum with the mean and the sum of squares with the variance).
A finite state machine for these semantics is shown in Figure~\ref{fig:state}.

One can not only add a vector of data points to an accumulator, but also a
\code{computed} object, which creates and adds the data to the accumulator.
The result and computed semantics taken together allow one to use ALEA with
very large random vectors which may fit in memory only once.

\paragraph{Reductions and serializations}
For parallel codes, all results support reduction (averaging over elements) through
the \code{reduce()} method, which takes an instance of the abstract
\code{reducer} interface. Depending on the implementation of the
reducer, the reduction is performed over different instances (threads,
processes, etc.) using MPI, OpenMP, ssh etc. without introducing
a hard dependency on any of these libraries.

Similarly, all results support serialization (converting to permanent
format) though the \code{serialize()} method and deserialization
through the \code{deserialize()} method, which take an instance
of the abstract \code{serializer} and \code{deserializer} interface,
respectively. Depending on the implementation, serialization to HDF5
as well as support for the HPX parallelization framework \citep{HPX} is provided.

\paragraph{Nonlinear propagation of uncertainty}
Transformations of results can be performed using the \code{transform} method.
The definition of \code{transform}, somewhat simplified, is:
\begin{cpplisting}
  template <typename Strategy, typename Function, typename Result>
  Result transform(Strategy strategy, Function f, const Result &x);
\end{cpplisting}
where \code{strategy} is a tag denoting the strategy for error propagation,
\code{f} is the function $f$, and \code{x} is the stochastic result.

Care has to be taken to correctly propagate the uncertainties if these
functions are non-linear, since neglecting the proper error propagation
leads to biasing both the mean and the variance of the transformed result.
Several strategies exist for performing these transforms, and we summarize the
ones supported by ALEA in Table~\ref{tab:prop}.  Similar to the case of the
accumulators, there is a tradeoff between computer time required (in particular
where $f$ has to be evaluated repeatedly) and requirements on the \code{Result}
type, such as the presence of binned data.\cite{Wallerberger18}

If $f$ is a linear function, propagation through \code{linear\_prop} is exact.
For non-linear functions on moderate number of components $k$, we suggest the
use of Jackknife (\code{jackknife\_prop}), which strikes a good tradeoff between
bias correction and cost.  The strategies \code{sampling\_prop} and
\code{bootstrap\_prop} will be implemented in a later version.

\begin{table}
\begin{centering}
\begin{tabular}{lllllcccc}
\toprule
\multirow{2}{*}{%
\begin{minipage}[t][1\height]{5em}%
propagation\\
strategy%
\end{minipage}} & \multirow{2}{*}{%
\begin{minipage}[t]{5em}%
preimage\\
distribution%
\end{minipage}} & \multirow{2}{*}{%
\begin{minipage}[t]{2.5em}%
mean\\
bias%
\end{minipage}} & \multirow{2}{*}{%
\begin{minipage}[t]{4em}%
variance\\
bias%
\end{minipage}} & \multirow{2}{*}{$N_{f}$} & \multicolumn{4}{c}{requires}\tabularnewline
\cmidrule{6-9}
 &  &  &  &  & $\langle x\rangle$ & $\Sigma_{x}$ & $\tau_{x}$ & $\bar{x}_{i}$\tabularnewline
\midrule
\code{no\_prop} & any & $N^{-1}$ & --- & $1$ & \checkmark &  &  & \tabularnewline
\code{linear\_prop} & Gauss & $N^{-1}$ & $1$ & $k$ & \checkmark & \checkmark & \checkmark & \tabularnewline
\code{sampling\_prop} & known & $\text{\ensuremath{S^{-\frac{1}{2}}}}$ & $S^{-\frac{1}{2}}$ & $S$ & \checkmark & \checkmark & \checkmark & \tabularnewline
\code{jackknife\_prop} & any & $B^{-2}$ & $B^{-2}$ & $B$ &  &  &  & \checkmark\tabularnewline
\code{bootstrap\_prop} & any & $S^{-\frac{1}{2}}$ & $S^{-\frac{1}{2}}$ & $S$ &  &  &  & \checkmark\tabularnewline
\bottomrule
\end{tabular}
\par\end{centering}
\caption{Common strategies for uncertainty propagation through a function $Y=f(X)$.
The preimage distribution is the distribution of $X$. Bias describes
the asymptotic $O(\cdots)$ worst-case behavior of the bias on the
mean in terms of the bin size $B$, the number of measurements $N$
taken, or the number of samples $S$ used for resampling. $N_{f}$
summarizes the number of function invocations required. The statistical
estimates required by each strategy are marked with ``\checkmark''. $\langle x\rangle$
is the sample mean, $\Sigma_{x}$ is the sample covariance matrix,
$\tau_{x}^{\mathrm{int}}$ is the integrated autocorrelation time,
and $\bar{x}_{i}$ are the bin means.}
\label{tab:prop}
\end{table}

\paragraph{Hypothesis testing}
A common task arising in statistical post-processing is to compare
a stochastic result with a deterministic one, e.g., when testing the
algorithm against known benchmark results, as well  as comparing two stochastic
results with each other, e.g., when trying to determine whether an
iterative procedure has converged. This is usually done by visual
inspection or some \emph{ad hoc} criterion.

Statistical hypothesis testing\citep{Wallerberger17} provides
a controlled alternative to these methods. ALEA provides the method
\code{test\_mean(x, y)}, which allows for hypothesis testing. If \code{x} is a statistical result
with proper (co-)variances and \code{y} is a benchmark (or vice-versa),
\code{test\_mean} performs Hotelling's $T^{2}$ test for the following null
hypothesis $H_0$:\cite{hotelling-amstat31}
\begin{equation}
H_{0}:\langle x\rangle=y,\label{eq:h0x}
\end{equation}
whereas if both are statistical results, the null hypothesis
\begin{equation}
H_{0}:\langle x\rangle=\langle y\rangle\label{eq:h0xy}
\end{equation}
is checked. In both cases, an object providing the test score and
the corresponding $p$-value is returned, which, roughly speaking, indicates
the probability of the two results being in agreement.  The $p$-value can
then be compared with a suitable threshold $\alpha\in[0,1)$, and
one considers the results equal (accepts the null hypothesis) if $p>\alpha$.

\section{Green's function library}
\label{sec:gf}

The \emph{Green's functions} component provides a type-safe interface to manipulate
objects representing bosonic or fermionic many-body Green's functions,
self-energies, susceptibilities, polarization functions, and similar
objects. From a programmer's perspective, these objects are
multidimensional arrays of floating-point or complex numbers, defined
over a set of meshes and addressable by a tuple of indices, each
belonging to a grid.  Currently, real frequency, Matsubara (imaginary frequency),
imaginary time (uniform, power,~\cite{Ku00} Legendre~\cite{Boehnke11} and 
Chebyshev~\cite{PhysRevB.98.075127}), momentum space, real space, and arbitrary 
index
meshes are supported.

\begin{figure}
\begin{centering}
\includegraphics[width=0.95\textwidth]{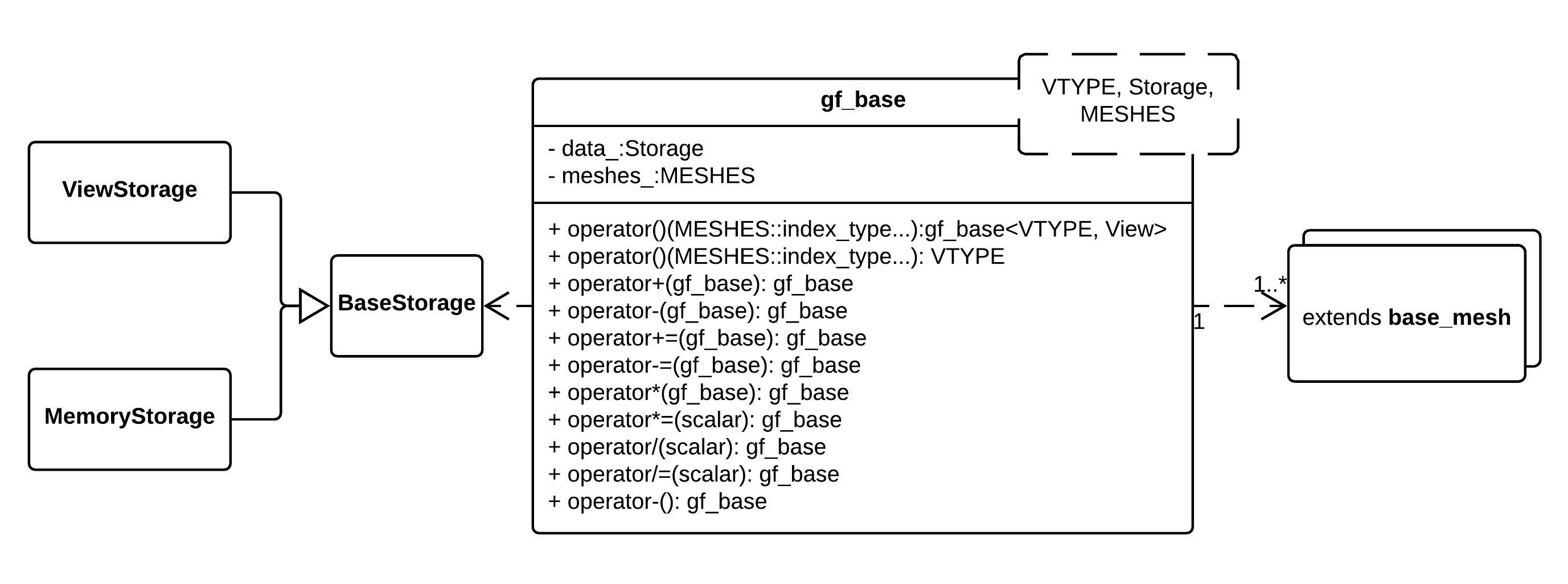}
\par\end{centering}
\caption{UML class diagram of Green's function library, indicating the main
classes in the Green's function library.  Each box corresponds to a class, and
arrows correspond to relationships between classes.}
\label{fig:uml-gf}
\end{figure}

These many-body objects often need to be supplemented with analytic tail information
encapsulating the high frequency / short time moments of the Green's
functions, so that high precision Fourier transforms, density
evaluations, or energy evaluations can be performed. In addition to this 
functionality, the Green's function component supports saving data to and 
loading it from binary \HDF files. The UML diagram illustrating
the Green's function classes and their interdependencies is given in 
Fig.~\ref{fig:uml-gf}.

The central class of the Green's function library is \code{gf\_base}. This 
class should be parametrized by the scalar type, the type of storage
and the list of meshes (grids). It provides the following basic arithmetic 
operations:
\begin{itemize}
  \item Addition/subtraction;
  \item In-place addition/subtraction;
  \item Multiplication by a constant.
\end{itemize}
The addition/subtraction operations can be performed on the Green's function 
objects that are defined on the same grid and have the same shape of the data 
array.
There are two possible storage types for this class. The memory storage version 
of the class is called \code{greenf}, whereas the view-based class is called 
\code{greenf\_view}. Instead of having only pre-defined Green's function 
classes of fixed dimension, we provide a flexible interface that makes it 
possible to declare Green's function types of any dimension. The following 
example defines 4-dimensional Green's function on Matsubara, arbitrary index, 
imaginary time and Legendre meshes:
\begin{cpplisting}
  // define meshes
  alps::gf::matsubara_positive_mesh x(100, 10);
  alps::gf::index_mesh y(10);
  alps::gf::itime_mesh z(100, 10);
  alps::gf::legendre_mesh w(100, 10);

  greenf<double, alps::gf::matsubara_positive_mesh, alps::gf::index_mesh, alps::gf::itime_mesh, alps::gf::legendre_mesh> g(x, y, z, w);
\end{cpplisting}

Using view-storage Green's function one can easily get access to a part of 
another Green's function object using index slicing. This idea is illustrated 
by the following code,
\begin{cpplisting}
  // define meshes and Green's function object 
  alps::gf::matsubara_positive_mesh x(100, 10);
  alps::gf::index_mesh y(20);
  greenf<double, alps::gf::matsubara_positive_mesh, alps::gf::index_mesh> g(x,y);
  
  // loop over leading index
  for(alps::gf::matsubara_positive_mesh::index_type w(0); w<x.extent(); ++w) {
    greenf_view<double, alps::gf::index_mesh> g2 = g(w);
    // do some operation on the Green's function view object
    // defined on the 'y' mesh alone
  }
\end{cpplisting}
In the current implementation we support only a slicing over a set of the
leading indices. More details on the Green's function library can be found
in the respective tutorial.

\section{Some of the other key library components}
\label{sec:components}

In this section, we provide a brief overview of the purpose and functionality of some of
the key components comprising the ALPS core libraries.  For a more detailed exposition
with a working example, see the previous publication\cite{ALPSCore17}.
Additional information is
available online and as part of the \Doxygen code documentation.
It should be noted that the components maintain minimal interdependence, and 
using a component in a program does not bring in other components,
unless they are required dependencies, in which case they will be used automatically.

\paragraph{Building a parallel Monte Carlo simulation} A generic Monte Carlo simulation
can be easily assembled from the classes provided by the \emph{Monte
  Carlo Scheduler} component. The programmer needs to define only the
problem-specific methods that are called at each Monte Carlo step,
such as methods to update the configuration of the Markov chain and to
collect the measured data. The simulation is parallelized
implicitly, using one \MPI process per
chain.

\paragraph{Storing, restoring, and checkpointing simulation results} To store the results of
a simulation in a cross-platform format for subsequent analysis, one can
use the \emph{Archive} component. The component provides convenient
interface to saving and loading of common \CXX data structures
(primitive types, complex numbers, STL vectors and maps), as well as
of objects of user-defined classes to/from \HDF~\cite{hdf5}, which is a
universally supported and
machine independent 
data format.

\paragraph{Reading command-line arguments and parameter files} Input
parameters to a simulation can be passed via a combination of a
parameter file and command line arguments. The \emph{Parameters}
library component is responsible for parsing the files and the command
line, and providing access to the data in the form of an
associative array (akin to \CXX \verb|map| or \Python dictionary). The
parameter files use the standard ``\verb|*.ini|'' format, a plain text
format with a line-based syntax containing \verb|key = value| pairs,
optionally divided into sections.

\section{Prerequisites and Installation}
\label{sec:prereqs}
To build the ALPS core libraries, any recent \CXX compiler can be
used; the libraries are tested with \GCC \cite{gcc} 4.8.1 and above,
Intel \cite{icc} \CXX 15.0 and above, and Clang \cite{clang} 3.2 and
above. The library follows the \CXX{}11
standard \cite{cpp11} to facilitate the portability to a
wide range of programming environments, including HPC clusters with older compilers. 
The library depends on the following external packages:
\begin{itemize}
\item The \CMake build system \cite{cmake} of version 3.1 and above.
\item The \Boost \CXX libraries \cite{boost} of version 1.56.0 and
  above. Only the headers of the Boost library are required. 
\item The \HDF library \cite{hdf5} version 1.8 and above.
\item The Eigen library \cite{Eigen3} version 3.3.4 and above (can be requested to be downloaded automatically).
\end{itemize}
To make use of (optional) parallel capabilities, an \MPI implementation
supporting standard 2.1~\cite{mpi-2.1} and above is required. Generating the developer's documentation requires \Doxygen
\cite{doxygen}  along with its dependencies.

The installation of the ALPS core libraries follows the standard procedure for any
\CMake-based package. The first step is to download the ALPS core libraries source
code; the recommended way is to download the latest ALPS core libraries release
from \url{https://github.com/ALPSCore/ALPSCore/releases}. Assuming
that all above-mentioned prerequisite
software is installed, the installation consists of unpacking the
release archive and running \CMake from a temporary build directory, as
outlined in the shell session example below (the \verb|$| sign designates a shell prompt):
\begin{bashlisting}[emph={tar,mkdir,cmake,make},emphstyle={\color{darkgreen}}]
$ tar -xzf ALPSCore-2.3.0.ta#()r.gz
$ mkdir build
$ cd build
$ export ALPSCore_DIR=$HOME/software/ALPSCore
$ cmake -DCMAKE_INSTALL_PREFIX=$ALPSCore_DIR \
        -DALPS_INSTALL_EIGEN=yes \
        ../ALPSCore-2.3.0 
$ make
$ make t#()est
$ make install
\end{bashlisting} 
The command at line~1 unpacks the release archive (version~2.3.0 in
this example); at line~4 the destination install directory of the
ALPS core libraries is set (\verb|$HOME/software/ALPSCore| in this
example). 
At line~6 the downloading and co-installation of Eigen library is
requested (see ~\ref{sec:installation-detail} for more details).

The ALPS core libraries come with an extensive set of tests;
it is strongly recommended to run the tests (via
\verb|make test|) to verify the
correctness of the build, as it is done at line~9 in the example
above.

The installation procedure is outlined in more details
in~\ref{sec:installation-detail}; Also, the file 
\verb|common/build/build.jenkins.sh| in the library release
source tree contains a build and installation script that can be further
consulted for various build options. 

Binary packages are available for some operating systems.
On macOS operating system, the ALPS core libraries package can be downloaded
and installed from the \code{\mbox{MacPorts}}~\cite{macports}
repository, using a command \verb|port install alpscore|.
On GNU/Debian Linux operating system, the ALPS core Debian package is provided by the MateriApps LIVE! project~\cite{malive} (see \cite{ma-alpscore} for more details).

\section{License and citation policy}
\label{sec:cite}
The GitHub version of ALPS core libraries is licensed under the GNU General Public
License version 2 (GPL~v.~2)~\cite{gplv2} or later.  The older ALPS license under which previous versions of the
code were licensed \cite{ALPS2.0} has been retired. We kindly request
that the present paper be cited, along with any relevant original
physics or algorithmic paper, in any published work utilizing an application code
that uses this library.

\section{Summary}
\label{sec:summary}
We have presented an updated and repackaged version of the core ALPS libraries, a lightweight \CXX library, designed to
facilitate rapid development of computational physics applications, 
and have described its main features.

The new version contains the next-generation statistics library ALEA, a major
overhaul of the Green's function library as well as updates to other components.

\section{Acknowledgments}
\label{sec:ackn}
Work on the ALPS library project is supported by the Simons collaboration on the many-electron problem. Aspects of the library are supported by NSF DMR-1606348 and DOE ER 46932.

\appendix

\section{Detailed installation procedure}
\label{sec:installation-detail}
In the following discussion we assume that all
prerequisite software (section~\ref{sec:prereqs}) is installed, and the ALPS core libraries
release (here, release~2.3.0) is downloaded into the current directory
as \verb|ALPSCore-2.3.0.tar.gz|. Also, we assume that the libraries are
to be installed in \verb|ALPSCore| subdirectory of the current user's
home directory. The commands are given assuming \verb|bash| as a user
shell. 

The first step is to unpack the release archive and set the desired
install directory:
\begin{bashlisting}[emph={tar,mkdir,cmake,make},emphstyle={\color{darkgreen}}]
$ tar -xzf ALPSCore-2.3.0.t#()ar.gz
$ export ALPSCore_DIR=$HOME/software/ALPSCore
\end{bashlisting} 

The next step is to perform the build of the library
(note that the build should not be performed in the source directory):
\begin{bashlisting}[emph={tar,mkdir,cmake,make},emphstyle={\color{darkgreen}},firstnumber=last]
$ mkdir build
$ cd build
$ cmake ../ALPSCore-2.3.0 -DCMAKE_INSTALL_PREFIX=$ALPSCore_DIR \
                          -DALPS_INSTALL_EIGEN=yes
\end{bashlisting}

The \verb|cmake| command at lines~5 and~6 accepts additional arguments
in the format \code{-D\textit{variable}=\textit{value}}. A number of
relevant \CMake variables is listed in Table~\ref{tab:cmake-args}. The
installation process is also affected by environment variables, some
of which are listed in Table~\ref{tab:cmake-env}; 
the \CMake variables take precedence over the environment variables.
The build and
installation script \verb|common/build/build.jenkins.sh| in the
ALPS core libraries release source tree provides an example of using some of the
build options.

It should be noted that line~6 of the example above will cause the
ALPSCore installation process to download a copy of Eigen
library\cite{Eigen3} and co-install it with ALPSCore. If the Eigen
library is already installed on your system, it may be preferable to
use the installed version. In this case, instead of requesting the installation of Eigen,
the location of the Eigen library should be specified:
\begin{bashlisting}[emph={tar,mkdir,cmake,make},emphstyle={\color{darkgreen}},firstnumber=5]
$ cmake ../ALPSCore-2.3.0 -DCMAKE_INSTALL_PREFIX=$ALPSCore_DIR \
                          -DEIGEN3_INCLUDE_DIR=/usr/l#()ocal/Eigen3
\end{bashlisting}
In this example, the location of the Eigen library is assumed to be
\code{/usr/local/Eigen3}; the actual location depends on your local
Eigen installation. The directory specified by the
\code{-DEIGEN3\_INCLUDE\_DIR} option must contain the \code{Eigen}
subdirectory.

\begin{table}
  \centering
  \begin{tabularx}{\textwidth}{lc>{\raggedright\arraybackslash}X}
    \textbf{Variable} & \textbf{Default value} & {\hfil\textbf{Comment}\hfil}\\
    \toprule
    \code{CMAKE\_CXX\_COMPILER} & (system default) & {Path to \CXX compiler executable.*} \\\midrule
    \code{ALPS\_CXX\_STD} & \code{c++11} & {The C++ standard to compile ALPSCore with.} \\\midrule\\
    \code{CMAKE\_INSTALL\_PREFIX} & \code{/usr/local} & {library target install directory.} \\\midrule
    \code{CMAKE\_BUILD\_TYPE} & RelWithDebInfo &  {Specifies build type.} \\\midrule
    \code{BOOST\_ROOT} &   & {\Boost install directory.
                            Set if \CMake fails to find \Boost.} \\\midrule
    \code{Boost\_NO\_SYSTEM\_PATHS} & \code{false} & {Set to \code{true} to disable search in default system directories,
                                           if the wrong version of \Boost is found.} \\\midrule
    \code{Boost\_NO\_BOOST\_CMAKE} & \code{false} & {Set to \code{true} to disable search for \Boost \CMake file,
                                          if the wrong version of \Boost is found.} \\\midrule
    \code{Documentation} & \code{ON} & Build developer's documentation. \\\midrule
    \code{ENABLE\_MPI} & \code{ON} & Enable \MPI build (set to \code{OFF} to disable). \\\midrule
    \code{Testing} & \code{ON} & Build unit tests (recommended). \\\midrule
    \code{ALPS\_BUILD\_TYPE} & \code{dynamic} & {Can be \code{dynamic} or
      \code{static}: build libraries as dynamic (``shared'') or static
      libraries, respectively.*}  \\\midrule
    \code{EIGEN3\_INCLUDE\_DIR} &  & Location of Eigen headers* (containing \code{Eigen} subdirectory) \\\midrule
    \code{Eigen3\_DIR} &  & Location of CMake-based Eigen installation* (containing \code{Eigen3Config.cmake}) \\\midrule
    \code{ALPS\_INSTALL\_EIGEN} & \code{NO} & set to \code{yes} to request downloading and co-installation of Eigen* \\\midrule
    \code{ALPS\_EIGEN\_UNPACK\_DIR} & (autodetected) & directory where to unpack Eigen for co-installation \\\midrule
    \code{ALPS\_EIGEN\_TGZ\_FILE} & (autodetected) & location of Eigen archive to unpack \\\midrule
    \bottomrule
    \multicolumn{3}{l}{{}*\footnotesize Note: For the change of this variable
        to take effect, remove your build directory and redo the build.}
  \end{tabularx}
  \caption{\CMake arguments relevant to building of ALPS core libraries.}
  \label{tab:cmake-args}
\end{table}

\begin{table}
  \centering
  \begin{tabularx}{1\textwidth}{l@{\hspace{5em}}X}
    \textbf{Variable} & {\hfil\textbf{Comment}\hfil}\\
    \toprule
    \code{CXX} &  Path to \CXX compiler executable.* \\\midrule
    \code{BOOST\_ROOT} &  \Boost install directory.
                        Set if \CMake fails to find \Boost. \\\midrule
    \code{HDF5\_ROOT}  &  \HDF install directory.
                        Set if \CMake fails to find \HDF.\\\midrule
    \code{Eigen3\_DIR}  &  Eigen install directory. Set if you use \CMake-based 
    installation of Eigen.\\

    \bottomrule
    \multicolumn{2}{l}{{}*\footnotesize Note: For the change of this variable
        to take effect, remove your build directory and redo the build.}
  \end{tabularx}
  \caption{Environment variables arguments relevant to building of ALPS core libraries.}
  \label{tab:cmake-env}
\end{table}

\end{document}